# Implications upon theory discrimination of an accurate measurement of the time rate of change of the gravitational "constant" *G* and other cosmological parameters

A.J. Sanders[1], G.T. Gillies[2], and E. Schmutzer[3,*]

[1] Department of Physics, University of Tennessee, Oak Ridge National Laboratory, MS 6054, Oak Ridge, TN 37831, USA
[2] Department of Physics, University of Virginia, Charlottesville, VA 22901, USA
[3] Theoretisch-Physikalisches Institut, Friedrich Schiller Universität, Max-Wien-Platz 1, 07743 Jena, Germany

[*] Corresponding author E-mail: ESchmu@aol.com

A substantial improvement in the accuracy of $\dot{G}$ tests (The dot denotes the time derivative.) would make it realistic to speak in terms of a *measurement* of $\dot{G}$, rather than merely a smaller upper bound on $|\dot{G}|$. We show that the accuracy $\Delta|\dot{G}/G| \approx 10^{-14}$ yr$^{-1}$ may be sufficient, given the accuracy of other cosmological parameters, to observe effects predicted by higher dimensions theories and, hence, to discriminate among different models. The $\dot{G}$ design goal for the SEE (Satellite Energy Exchange) mission is $\Delta(\dot{G}/G) \approx 10^{-14}$ yr$^{-1}$.

## 1 Introduction and different theoretical approaches to $\dot{G}$

We attempt to show here that a precise measurement of $\dot{G}$, *i.e.*, the rate of change of the gravitational "constant" *G*, would be very significant for testing post-Einsteinian theories, given the accuracy that is now available or will likely soon be available for other cosmological parameters.

A striking feature of most post-Einsteinian theories is that they cannot retain a constant *G*, but rather require various secular rates of change; the assumption that the "coupling constant" *G* is actually constant is not consistent with these unification theories. Rather, most of the current promising approaches to unification theory, including string theories, brane theories, and supergravity, incorporate the gravitational force at a fundamental level. Moreover, it has long been recognized that a measurement of $\dot{G}$ can provide one of the very few experimental "windows" for discriminating among various multidimensional theories (see, for example, Marciano [1], Damour, Gibbons and Taylor [2], Ivashchuk and Melnikov [3,4], Bronnikov, Ivashchuk and Melnikov [5], Melnikov [6,7], and Drinkwater *et al*. [8]).

The need to test theoretical models has acquired new impetus with the publication in 2006 of two works decrying the lack of experimental tests of string theories, by Woit [9] and Smolin [10].

More broadly, gravitation is perhaps the most important missing link in efforts to achieve a satisfactory unified theory of physics. As pointed out by Zichichi [11], our understanding of gravitation remains incomplete in contrast with that of the other known forces of Nature - strong, weak, and electromagnetic - which have been explored in great detail during the past six decades and are now understood with extraordinary precision. However, the relative lack of precise experimental evidence at present on a number of open questions in gravitational physics makes it very difficult to assess the validity of alternative schemes that have been proposed.

In addition, the discovery that the expansion of the Universe is accelerating [12,13,14,15] - rather than decelerating, as previously expected - has raised further fundamental questions about the structure of space-time and about the possible existence of some form of "Dark Energy," or "quintessence" [16,17,18]. The need for experimental windows on these questions - and, hence, on cosmology and unification theory - is another very strong motivating factor for a precise measurement of $\dot{G}$.

Perhaps most importantly, an unequivocal experimental demonstration that a fundamental "constant" of Nature is not constant would be of significant scientific value in any case, even if it did not immediately lead to a decisive discrimination among existing theories. It would invigorate interest in the new theories, most of which do in fact predict time variation of $G$ and other fundamental "constants." A finding of non-zero $\dot{G}$ would at the very least require revision of general relativity, since it assumes a constant value of $G$. More broadly, such a finding would clearly mark the boundaries where general relativity is valid, and specify the onset of new physics.

We examine here what can be learned from a precise measurement of $\dot{G}$ - either alone or in concert which various cosmological parameters - regarding the validity of theoretical models. We illustrate the possible findings using several different theoretical models. However, these models *per se* are not our focus; we do not seek to either validate or disprove any particular model. Rather, our purpose is to illustrate the capability of precise measurements to discriminate among different models.

Very significantly, we find that the numerical results from various models may differ by multiple orders of magnitude, although their equations may look very similar. A key implication of this finding is that it would be useful to have the calculations carried all the way to numerical results so that the *quantitative* significance of the models can be fully appreciated.

We emphasized above the experimental importance of whether $G$ is a true constant of nature or only an important parameter which arises in gravitational theories in a model-dependent way. In a number of modern theories gravitation arises from a true intrinsic constant, from which the observed "constant" $G$ arises. The specific form of the connection between the observed (variable) $G$ and the intrinsic constant is a distinguishing feature of various theoretical models. It seems to be probable that a consensus is emerging that the present array of models is likely too narrow to bridge the gap from the constant-$G$ concept of Newton and Einstein to a variable-$G$ that can satisfactorily accord with the high-precision data expected in the future.

Roughly speaking we need to look at two different approaches**:**

- Models that retain the four-dimensionality of space-time and accept an additional new field, introduced in the generalized Einsteinian field equations. This way automatically leads to the introduction of constant parameters with numerical values to be chosen.

- Models that employ a higher-dimensional (> 4) geometrical basis. This means a five-fold (Kaluza-Klein platform or projective relativity platform) or even higher-dimensionality (string and brane theories).

## 2   Present experimental status of $\dot{G}$

There has never been a laboratory measurement of $\dot{G}$ (using test masses in a controlled situation) at cosmologically interesting levels of precision. The extreme difficulty of such an undertaking has prevented even the best of the terrestrial experiments from reaching the presently foreseen requirements for testing theories. The observational - *vis-à-vis* experimental - evidence regarding $\dot{G}$ reveals scattered results; see Reasenberg [19], Gillies [20], and Will [21].

Lunar Laser Ranging (LLR) provides the present experimental bound on $\dot{G}$; see Williams *et al.* [22,23,24,25,26]:

$$\Delta|\dot{G}/G| \approx 2 \times 10^{-12} \text{ yr}^{-1}. \tag{1}$$

Kaspi, Taylor and Ryba [27] have proposed that binary pulsar systems could be interesting. But note that Will [21] states that binary pulsar results cannot be trusted because the mass varies, which makes the "experimental" result for $|\dot{G}/G|$ model-dependent. Further discussion on this point is in Benvenuto *et al.* [28].

Modern theories are expected to make predictions in the range of

$$\approx 10^{-13} \text{ yr}^{-1} < \Delta|\dot{G}/G| \lesssim 10^{-11} \text{ yr}^{-1} \tag{2}$$

or somewhat less.

The planned Satellite Energy Exchange (SEE) mission proposed originally by Sanders and Deeds [29,30,31] and discussed further by several others [32,33,34,35,36,37,38,39,40,41,42,43,44] has a design goal of measuring $\Delta|\dot{G}/G|$ to within $\approx 10^{-14}$ yr$^{-1}$.

Finally, we note that there is some discussion about whether or not the value of $(d\alpha/dt)/\alpha$ (see, *e.g.*, Prestage *et al.* [45]) is tied to that of $\dot{G}/G$, and hence that there is no need to test one if the other is known to be zero within tight limits ($\alpha$ is Sommerfeld's fine structure constant). However, such potential connections or lack thereof are model dependent (Moffat [46], Moffat and Gillies [47]).

## 3   Connections between theories and the values of cosmological constants

Different theoretical models may predict different values of $\dot{G}$, and the predicted value of $\dot{G}$ may be closely linked with other parameters of cosmological interest, such as the Hubble constant $H$, the "deceleration" parameter $q$, and the scalar-tensor parameter $\omega$. The types and strengths of the linkages are very model-dependent, which is fortuitous from the standpoint of providing ways to discriminate among models. In what follows we give numerical results for the predictions of $\dot{G}$, including the linkages to other variables, for several different models, and we explore especially how an accurate measurement of $\dot{G}$ might be a very sensitive indicator of a model's validity.

We reiterate that it is not our purpose to indicate the extent to which any particular model can be either validated or refuted by presently available experimental data. Rather, our treatment is intended to be illustrative. We believe that the climate for rigorous systematic testing of theoretical models is now healthier than ever before and that it will sharply improve in the near future because of two fundamental changes: first, the quality and precision of a variety of cosmological parameters are now increasing rapidly, as mentioned by Turner in the context of his views on the new era of "precision cosmology" [48], and second, the theoretical discussions can now provide considerably more quantitative material.

Thus, our approach anticipates the situation that may exist in the near future. We are especially interested in two things: the extent to which an accurate measurement of $\dot{G}$ may have the potential to be decisive in choosing among different theoretical models and how the value of $\dot{G}$ as a theory discriminator may be increased by future improvements in other cosmological variables. It is inevitable that the validity of some models will probably be challenged by the data that will become available during the next few years, and we anticipate that the comparisons of the models which we give here are representative of this process.

## 4  Calculations of secular change in *G* using various models

We first consider a model by Melnikov and colleagues [49] which uses 10-dimensional supergravity (p-brane theory requires 10 or 11 dimensions; see also Overduin and Wesson [50].). They find that a 10-dimensional cosmology with two-component anisotropic fluid and positive acceleration (negative "deceleration parameter" $q = -[1 + (dH/dt)/H^2]$, which describes the rate of deceleration or acceleration of the expansion of the Universe) leads to

$$\dot{G}/G \cong -4 \times 10^{-12} \text{ yr}^{-1}. \tag{3}$$

We note that this result is somewhat above the present experimental bound.

Observational data indicate that *q* is negative, *i.e.*, the expansion rate is increasing rather than decreasing, as had previously been expected. Moreover, the negative value of *q* implies that the age of the Universe is larger than *1/H* (A zero value would imply that the age of the Universe is simply *1/H*.).

Melnikov and colleagues [49] have also explored a second class of models, namely scalar-tensor (S-T) cosmologies, of which Jordan-Brans-Dicke theory is a special case. They have elucidated how $\dot{G}$ would be linked, in the context of these models, to the Hubble parameter *H*, to the S-T parameter ω, and to the deceleration parameter *q* through the shifted deceleration paramter $q´ \cong 1.1+2q$.

They focus on the special case in which the universe is spatially flat (so that the density ρ is the critical density $\rho_{cr}$). The expression from which $\dot{G}$ is calculated in this case is given by Eq. (4):

$$\omega g^2 + g - q´ = 0. \tag{4}$$

Here *g* is the ratio of $\dot{G}/G$ to the Hubble rate *H,* which Dirac [51,52] naturally assumed to be unity in his pioneering papers on his Large Numbers Hypothesis and $\dot{G}$. All of the $\dot{G}$ predictions in this section assume that the Hubble rate is approximately $H \cong 7 \times 10^{-11}$ yr$^{-1} \cong 70$ km s$^{-1}$ Mpc$^{-1}$.

Values of $\dot{G}/G$ obtained from Eq. (4) are shown in Fig. 1. We note that the values are within the experimental bound for a wide range of parameter values. Eq. (4) shows that *g* depends on the S-T parameter ω and on the deceleration parameter *q* through the shifted parameter $q´ \cong 1.1+2q$. The term 1.1 is the result of a calculation $1.1=3(1–x)–1$ used in deriving Eq. (4), where *x*=0.3 is taken to be the fraction of the total density which is comprised of ordinary matter (including dark matter). We note that the recent results from the NASA's Wilkinson Microwave Anisotropy Probe (WMAP) observatory suggest that $q´\cong 1.2+2q$ would be somewhat more accurate, since WMAP found that about 27% of the total density is due to ordinary matter. A key question in modern cosmology is the nature of the so-called "dark energy" and the field resulting from it which could cause this acceleration.

The S-T parameter ω is very large (ω>>2500) because it is the ratio of the tensor to the scalar amplitude, and general relativity is so nearly correct. (If the metric were pure tensor, as in general relativity, ω would be infinite.)

If we choose $\omega \gg 2500$ and $q' = 0$, we see from Eq. (4) and Fig. 1 that according to this theory

$$|\dot{G}/G| < 3 \times 10^{-14} \text{ yr}^{-1} \tag{5}$$

However, if $q'$ is chosen equal to the largest value consistent with observations, *viz*. $q' \cong 0.4$, then Eq. (4) and Fig. 1 allow a larger value of $\dot{G}$:

$$|\dot{G}/G| < 0.9 \times 10^{-12} \text{ yr}^{-1}. \tag{6}$$

We note that the error bars for a measurement of $\dot{G}/G$ at the level of $10^{-14}$ would be about 1 mm high on the scale of Fig. 1. Thus, it is clear that such accuracy, coupled with likely future improved knowledge of the deceleration parameter, would place extremely severe constraints on $\omega$.

A number of other authors have also investigated the relationships among $\dot{G}$ and the various parameters of higher-dimensions models. Chiba [53] as well as Perrotta, Baccigalupi and Matarrese [54] explored variants of Jordan-Brans-Dicke theory which they call "Extended Quintessence." The work of Perrotta and colleagues is based in part on that of Chiba.

For illustration these authors obtained numerical results for simple forms of the gravitational part of the Lagrangian. Chiba's choice of the gravitational part of the Lagrangian was:

$$R/2 \times [1/(8\pi G_{bare}) - \xi \phi^2] \tag{7}$$

See Chiba's Eq. (2.1) and the discussion in the immediately following paragraph. Chiba uses two different forms of the potential, namely an inverse-power law (his Eq. (2.6)) and an exponential (his page 2, 2nd column, near bottom), following Zlatev *et al*. [55] and Steinhardt *et al*. [56]. With these choices of potential, Chiba finds that the bound on $\xi$ inferred from the current experimental bound on $\dot{G}$ is

$$-10^{-2} \leq \xi \leq (10^{-2} \text{ to } 10^{-1}) \tag{8}$$

(Chiba's equation (2.23)). Here $\xi$ is a coupling parameter which indicates the degree of deviation from general relativity. Choosing $\xi = 0$ gives ordinary general relativity. Non-zero values of $\xi$ give various violations of GR, including non-zero $\dot{G}$ and non-infinite $\omega_{JBD}$. We note that Eq. (8) is roughly equivalent to the result obtained by Perrotta *et al*. in Eq. (13), when account is taken of the factor of $-2$ between the meanings of $\xi$ in the two papers.

An alternative test suggested by Perrotta *et al*. relies on the variation of the inverse Hubble length $(cH)^{-1}$ with the distance parameter $z$. From their Fig. 2, one may infer that the Hubble length at $z+1=1000$ is $\approx 26\%$ higher if $\xi$ is zero than if it has the illustrative value $10^{-2}$. It follows that, when the state of the art of Hubble measurements is capable of an error $\approx 1\%$, its variation with $z$ will constitute a measurement of $\xi$ with error $\approx 10^{-3}$. Although this is weaker than the present bound of $\xi$ from $\omega$ and the expected error on $\xi$ from $\dot{G}$, in the future Hubble measurements could possibly provide the necessary complement to other measurements to carry out a cosmologically-significant test of the "Induced Gravity" model or other similar models.

Perrotta *et al*. used two forms of the gravitational part of the Lagrangian, namely the Induced Gravity part (IG part)

$$F(\phi)R = \xi \phi^2 R, \tag{9}$$

as given in their Eq. (27), and the Nonminimal Coupling part (NMC part)

$$F(\phi)R = (1+\xi\phi^2)R$$
$$\text{or } (1/(8\pi G) + \xi\phi^2)R, \tag{10}$$

as given in their Eq. (30) and (31). (Note that the NMC form is essentially the same as Chiba's form (his Eq. (2.1)), but with $-\xi/2$ in Chiba's notation equivalent to $\xi$ in the notation of Perrotta *et al*.) Here $R$ is the Ricci curvature scalar and $\phi$ is the quintessence field.

The authors investigated the experimental bounds on $\xi$ and concluded that, at present, the tightest bounds come from the present lower bound on the Jordan-Brans-Dicke parameter - which the authors conservatively took as $\omega \geq 500$ - rather than from upper bounds on $\dot{G}$ obtained from Hellings *et al*. [57] and by Damour and Taylor [58], viz. $|\dot{G}/G| \approx 10^{-11}$ yr$^{-1}$. To wit, Perrotta *et al*. find that the bound on $\xi$ obtained from $\dot{G}$ is roughly

$$\xi \leq 3 \times 10^{-2} \quad \text{(IG and NMC)} \tag{11}$$

while the bounds obtained from $\omega$ are much tighter, namely roughly

$$\xi \leq 5 \times 10^{-4} \quad \text{(IG)} \tag{12}$$

and

$$\xi \leq 5 \times 10^{-3} \quad \text{(NMC)}. \tag{13}$$

We note that all these bounds on $\xi$ might be tightened considerably by using more recent bounds on $\omega$ and $\dot{G}$, but nevertheless $\omega$ would be more restrictive than $\dot{G}$.

In the NMC case, Perrotta *et al*. arrived at their conclusions by appealing to a definition of the Planck mass ($m_{Pl}$) that sets it equal to $(\hbar c/G)^{1/2}$, which in turn is taken to be the low-energy value of the quintessence coupling parameter. By choosing this particular parameterization, they make it possible to interpret predictions of the quintessence field strength in terms of the quantity $m_{Pl}$ which is rather well understood in relativistic theories of gravity and cosmology.

Perrotta *et al*. show the variation of $(\dot{G}/G)$ with $\xi$ explicitly in their Fig. 1, which is obtained by numerical integration of their differential equations. From this figure, we have inferred that the approximate relationships, are, for the IG model,

$$\dot{G}/G = -3.44 \times 10^{-10} \xi \tag{14}$$

and, for the NMC model,

$$\dot{G}/G = -7.24 \times 10^{-9} \xi^2 - 5.61 \times 10^{-11} \xi \tag{15}$$

in units of yr$^{-1}$. Eq. (14) and (15) are plotted in our Fig. 2. We note that in the NMC case the relationship is nearly linear for very small $\xi$.

In both these models $\omega$ is also a function of the coupling parameter $\xi$. We infer that it follows that $\dot{G}$ and $\omega$ *must be implicit functions of each other* in the context of these models. Accordingly, we have made this relationship explicit.

Our results are shown in Fig. 3. For comparison, we have also re-plotted in the same figure the results of Chiba's model and Melnikov's S-T models from Eq. (4) of Fig. 1. In Fig. 1 the shifted deceleration parameter $q'$ was shown as a continuous variable, and only two discrete values of $\omega$ were chosen (thus reversing of the procedure of Fig. 3).

The error bars in the vertical direction in Figure 3 represent the error in the theoretical prediction of $\omega$ if $\dot{G}/G$ were known to within $1 \times 10^{-14}$ yr$^{-1}$.

The lower portion of Figure 3 ($\omega \approx 10^3$) is obviously not physically allowed because $\omega$ is already known to be large. In the physically-allowed region of Figure 3 it is evident that a combination of an accurate measurement of $\dot{G}$ and a rudimentary measurement of ω (say, to within an order of magnitude) would discriminate strongly against at least several of these models. For example, the vertical dotted lines in Figure 3 indicate the situation, in the case of an experimental measurement where $\dot{G}/G = (5 \pm 1) \times 10^{-14}$ yr$^{-1}$. In this case, everything outside this narrow vertical strip would be unphysical. Although this fact alone would not exclude any of the models illustrated in the figure, combining such a measurement of $\dot{G}/G$ with a test of $\omega_{JBD}$ could greatly constrain the physically-allowed region; *e.g.*, if

$$10^4 < \omega_{JBD} < 10^5 . \tag{16}$$

Thus, perhaps only one model (Melnikov with $q' \cong 0.1$) among those illustrated would be compatible with observation. More broadly, if $\dot{G}/G$ were measured to be anywhere in the middle of the range shown (*i.e.*, $\dot{G}/G \approx 10^{-13}$ yr$^{-1}$), then most of the corresponding values of ω for the different models would be separated by factors of $\approx 5$ to $\approx 10$, making it likely that many of the models would not survive such a test.

Let us as another example also mention the model of Miyazaki [59], who considers the Machian cosmological solution which satisfies $\phi = O(\rho/\omega)$ for the perfect fluid with negative pressure. When the coefficient of the equation of state γ approaches $-1/3$, the gravitational parameter approaches a truly constant value. If the present density is assumed to be $\rho_o \approx \rho_{cr}$ (critical density), the parameter ε (where $\gamma = (\varepsilon - 1)/3$) must have a value of order $10^{-3}$ to support the present value of the gravitational constant. The closed model is valid for $\omega < -3/2\varepsilon$ and exhibits a slowly accelerating expansion, in agreement with observation. All of this makes it possible to understand why Miyazaki's coupling parameter |ω| is so large ($\omega \approx -10^3$). The time-variation of the gravitational constant at the present time

$$|\dot{G}/G| \approx 10^{-13} \text{ yr}^{-1} \tag{17}$$

is derived in this model.

As a further remark to this theme we recall that among the early work with explicit numerical results are the predictions of Wu and colleagues, who found that

$$\dot{G}/G = (q_o - 13\Omega_o H_o^2 t_o^2/8)/t_o \tag{18}$$

holds, where $q_o$ is the deceleration parameter, $\Omega_o$ is the critical-density ratio, $H$ is the Hubble parameter, and $t_o$ is the age of the universe [60,61,62]. Evaluating Eq. (18) gives the result $\dot{G}/G \approx -10^{-10}$ yr$^{-1}$, which is inconsistent with present experimental findings.

In context with the application of the Scalar-Tensor Gravity to the Earth-Moon dynamics, using the optical interferometry of Lunar Laser Ranging (LLR), we would like to emphasize the contributions of Nordtvedt [63,64,65].

Our final remark refers to the time-dependence of other constants of Nature: For more than two decades several investigators have explored the possibility of time variation of other fundamental "constants" of Nature. Among those hypothetically most plausible time-dependent quantities are: Electric elementary charge, Planckian elementary quantum of action, Planckian elementary length, Sommerfeld´s (dimensionless) interaction constant, Fermi constant of weak interaction, *etc*. Of course, most of these constants are in mutual relationship. Therefore, the finding of a time variation of one of them would lead to an explosion of basic changes in our picture of Nature and also shed new light on recent advances in cosmology. It is not possible here to cite all the recent papers of this field.

## 5  Projective Unified Field Theory

Another example of a theoretical model is the Projective Unified Field Theory (*Projektive Einheitliche Feldtheorie* or PUFT), which is a five-dimensional theory developed by one of the authors (ES). Broadly speaking this theory is a further development of the 5-dimensional projective relativity theories of a series of authors before World War II, including the mathematicians Veblen and Hoffmann [66], as well as van Dantzig and Schouten [67,68]. Jordan [69] should be mentioned in connection with the new insights provided by the additional scalar field, since he had some degree of success concerning the application of his concepts to geophysics based on Dirac´s hypothesis of a temporal change of the gravitational constant. However, this 5-dimensional approach should not be confused with Jordan´s 4-dimensional tensor-scalar theory [70] later elaborated further by Brans and Dicke [71], *viz*., the Jordan-Brans-Dicke scalar-tensor theory mentioned above.

The approach of Schmutzer starts with completely new geometric axioms for the five-dimensional projective space and is distinguished from previous work by an intrinsically succinct method for projecting 5-vectors onto the ordinary 4-dimensional space-time with Riemannian geometry. The field equations, balance equations, conservation laws and the physical interpretation of PUFT have evolved over the years, from its origin [72,73,74] in different stages [75,76,77,78] to the present [79, 80, 81,82,83,84]. The ingredients in this space-time are: gravity, electromagnetism and scalarism (as physical reality on the same physical level). The phenomenon scalarism/scalarity is described by the scalaric field function, being an outcome of the five-dimensionality similar to Jordan's scalar field mentioned above. The adjective to scalarism is "scalaric".

For all versions of PUFT, which have been evolving sequentially, the basis is a postulated 5-dimensional field equation with Einstein-like aspects, containing a variable cosmological term, from which, in the 4-dimensional space-time, a coupled non-linear system of second-order differential equations is obtained. An important term in this system is a definite 4-dimensional "scalaric cosmological term" with a dimensionless "scalaric cosmological constant" $\lambda_s$, which plays a role somewhat analogous to Einstein's cosmological constant.

In the cosmological model with homogeneity, isotropy and sperical symmetry, three time-dependent field functions (curvature radius of the cosmos, scalaric world function and mass density) occur. Solution of this system leads to the inclusion of matter conventionally, by the addition of a 5-dimensional energy projector $\Theta_{\mu\nu}$, which is multiplied by the usual "Einsteinian gravitational constant" $\kappa_o$. The two constants of Nature in PUFT are the Einsteinian gravitational constant $\kappa_o$ and the scalaric cosmological constant $\lambda_s$.

Since the 4-dimensional differential equations resulting by projection from PUFT need five initial conditions, these have been adjusted in order to obtain good agreement between the theoretical solution (numerically obtained) and the experimental results from WMAP.

The PUFT predictions are given for recent values of key cosmological observables [79]. Thus constrained to agree with WMAP and further experimental results, the following list shows the theoretical outcome for the present epoch:

1. Hubble parameter:

$$H = 70.96 \text{ km s}^{-1} \text{ Mpc}^{-1} \qquad (19)$$

2. Predicted value of $\dot{G}/G$:

$$\dot{G}/G \cong -1.49 \times 10^{-13} \text{ yr}^{-1} \qquad (20)$$

This agrees with the current experimental limit from LLR observations by Williams [25], *viz*.

$$|\dot{G}/G| < 1.8 \times 10^{-12} \text{ yr}^{-1} \qquad (21)$$

3. Deceleration parameter:

$$q = -1.02 \text{ (dimensionless)} \qquad (22)$$

Further predictions by PUFT of other key variables include:

4. Scalaric cosmological constant:

$$\lambda_s = 4.4 \times 10^{-122} \text{ (dimensionless)} \qquad (23)$$

5. Matter density:

$$\mu = 3.3 \times 10^{-33} \text{ kg m}^{-3} \qquad (24)$$

6. Age of cosmos:

$$t_{AC} = 13.7 \times 10^9 \text{ yr} \qquad (25)$$

7. Curvature radius of the cosmos:

$$R_{cosmos} = 3.5 \times 10^{26} \text{ m} \qquad (26)$$

A distinctive aspect of PUFT is that its picture of the evolution of the cosmos begins with a singularity-free "Big Start" (Urstart) rather than a singular "Big Bang" (Urknall). Further we should mention that the theoretical *finale* of the cosmos occurs at

$$t_{FC} = 27 \times 10^9 \text{ yr}. \qquad (27)$$

Finally we state further predictions of PUFT:

- This theory presents the numerically accepted value for the empirically confirmed effect of the switch from decelerated to accelerated expansion about $6 \times 10^9$ years ago. We note that this inherently avoids the requirement for dark energy to explain the observed accelerated expansion.
- This theory explains the well-known Einstein effects of general-relativistic motion (anomalous precession of the perihelion of Mercury, deflection of electromagnetic waves by a central mass, frequency shift of electromagnetic waves in a gravitational field, *etc.*).

We note that the numerical predictions of PUFT, mentioned above, are tightly interconnecteted. If for empirical reasons one of them had to be changed, the others would unavoidably change as well. Therefore, it is appealing at a fundamental level, which is important as we anticipate new and better (particularly) satellite experiments to improve the numerical values of PUFT listed above.

We further note that the Hubble parameter has a history of large empirical changes and uncertainties in its measured values, and there is some evidence that the value may be dependent on the experimental method. To wit, in its beginning [85] the gravitational lensing method yielded a significantly lower value,

$$H = (48 \pm 3) \text{ km s}^{-1} \text{ Mpc}^{-1} \qquad (28)$$

than the contemporaneous values obtained from the traditional method based on redshift-vs-distance, although a very recent re-analysis of gravitational-lensing data seems to reconcile the values obtained by the two methods [86]. (In passing we note that the distance measurements had also yielded much lower values of *H* until recently.) These differences and changes have potentially large cosmological significance, so this situation bears close watching in future.

## 6   Expected significance of a SEE mission

The general objective of the SEE Mission is to significantly advance the original experimental base [87,88] required for testing unification theories. As such, SEE is being designed to substantially increase the precision and accuracy of measurement of $\dot{G}$, since it is likely to be decisive in discriminating among modern theories. A SEE mission may hold promise of providing a *controlled* experiment for $\dot{G}$ measurement with test masses which, because they are free-floating in a low-disturbance environment in space, have the potential to provide very fine accuracy [29,33,34,35,36,37,38,41,42,43,89].

The predicted result from a SEE mission is rather tight, *viz.*

$$\Delta|\dot{G}/G| \approx 1\times10^{-14} \text{ yr}^{-1}. \tag{29}$$

Such potential accuracy would likely be sufficient to provide a *measurement*, not just an upper bound, which would significantly enhance the capability of discriminating among various possible unified theories. From the spread among their various predictions for $\dot{G}$, we see that the value of $\dot{G}$ can be a very sensitive probe of the type of cosmology and the values of key parameters. It is highly unlikely that, by coincidence, more than one theoretical model will predict any given value of $\dot{G}/G$ to within the accuracy expected from a SEE mission, *viz.* $\Delta|\dot{G}/G| \approx 10^{-14}$ yr$^{-1}$, while being consistent with the values of other cosmological parameters, such as $\omega_{JBD}$ (Fig. 3). Thus, a measurement of $\dot{G}/G$ at this accuracy may be able to discriminate strongly against all but a few specific theoretical models, thus performing a filtering role. The very precise experimental data which would be sought via a SEE mission could thus be key for advancing the prospects of unification theory.

## 7  Conclusion

It is well accepted that gravitation is the missing link in unification theory. In the new era of precision cosmology, it is reasonable to expect that very accurate observational data will increasingly provide guidance to theory by narrowing down both the list of physically-acceptable models and values of parameters in the remaining models. Different models can give very different numerical results, even if their functional structures appear similar. Therefore carrying theoretical predictions through to numerical estimates can be very helpful and, indeed in many cases, absolutely essential to the mutual support of theory and experiment. Combinations of variables tend to be more effective than individual variables for discriminating among models, and within that context, $\dot{G}/G$ has the potential to be a very effective window/discriminator, especially in combination with the scalar-tensor parameter ω.

### Acknowledgements

The authors presented an early version of this paper and the scientific challenges of a $\dot{G}$ measurement at meetings at the Peter Bergmann Center in Erice (Sicily, Italy). We are pleased to acknowledge the important role played by Prof. Venzo de Sabbata (deceased) and the Erice conferences that he organized, in helping to develop our collaboration. Further, they remember gratefully the visits of Prof. A. Zichichi, whose words [11] stimulated our thinking: "Although the theory of general relativity was formulated more than 80 years ago, gravitational forces are only now [2000] entering the arena for basic scientific research." We also thank Profs. Ken Nordtvedt, T. Chiba, Carlos Baccigalupi, Vitaly Melnikov, Kirril Bronnikov, and Voladya Ivashchuk for insightful suggestions. This work was supported largely by NASA grant NAG 8-1442 in the Fundamental Physics in Microgravity Program, and was discussed at the NASA "Quantum to Cosmos III" Conference at Airlie Center in July, 2008. We are also pleased to acknowledge support from a NATO Linkage Grant, an NSF travel grant, and a Scholarly Achievement and Research Incentive Grant from The University of Tennessee.

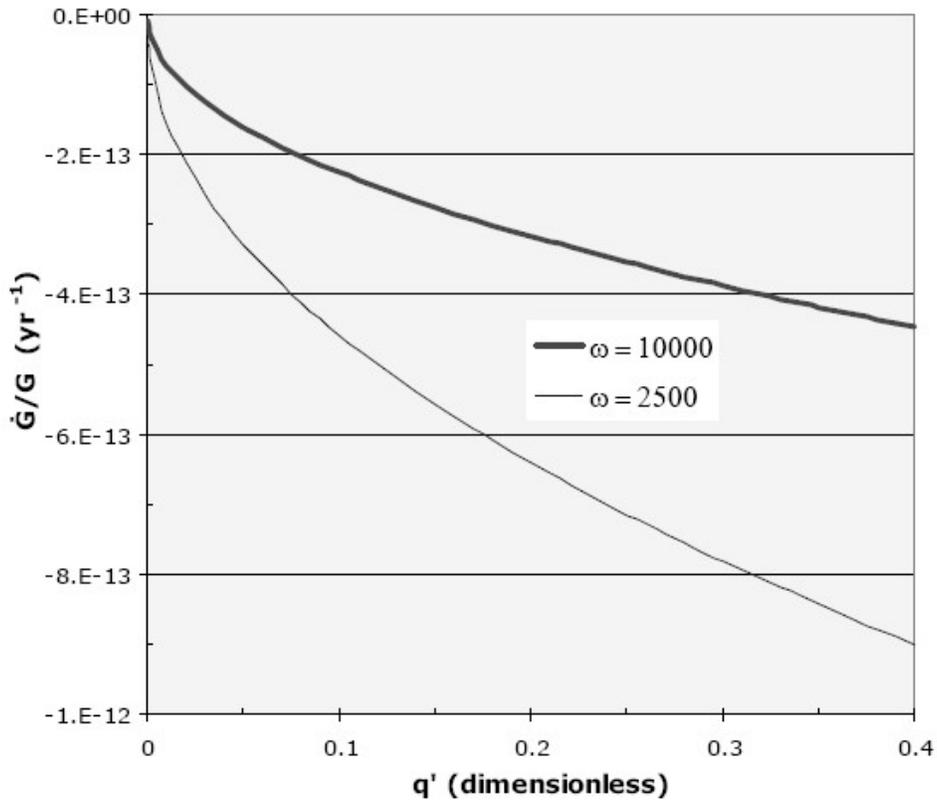

**Fig. 1** Ġ depends on the rate of acceleration of the expansion of the Universe in scalar-tensor theories.

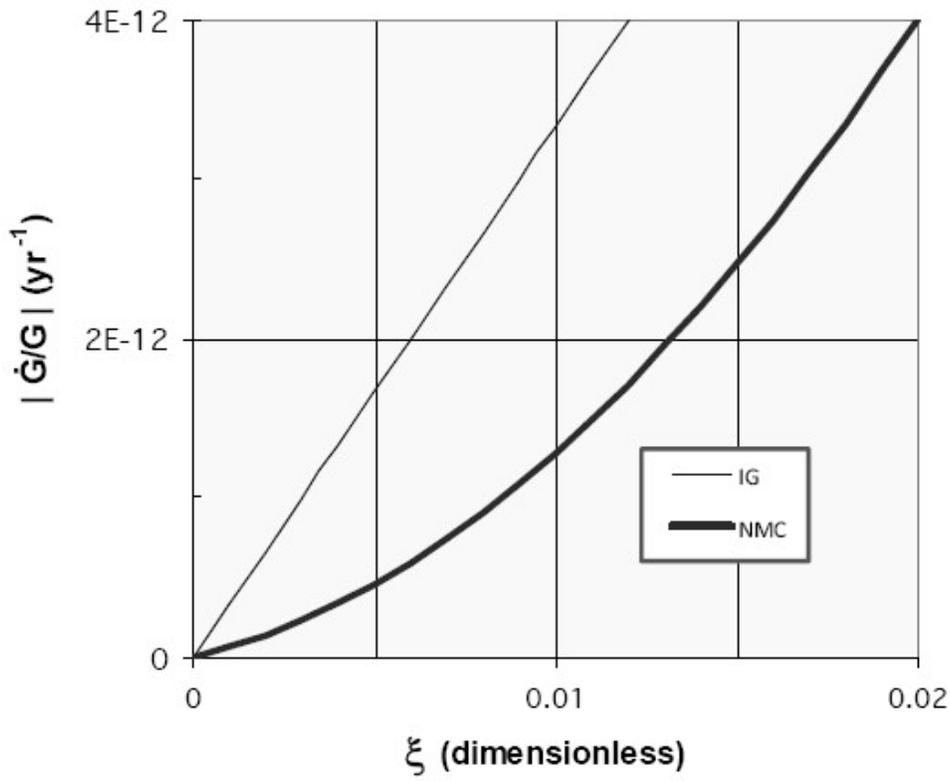

**Fig. 2** In the induced-gravity (IG) model of Perotta et al |Ġ/G| varies linearly with the hidden internal parameter $\xi$. In their non-minimal coupling (NMG) model, a quadratic term is also present.

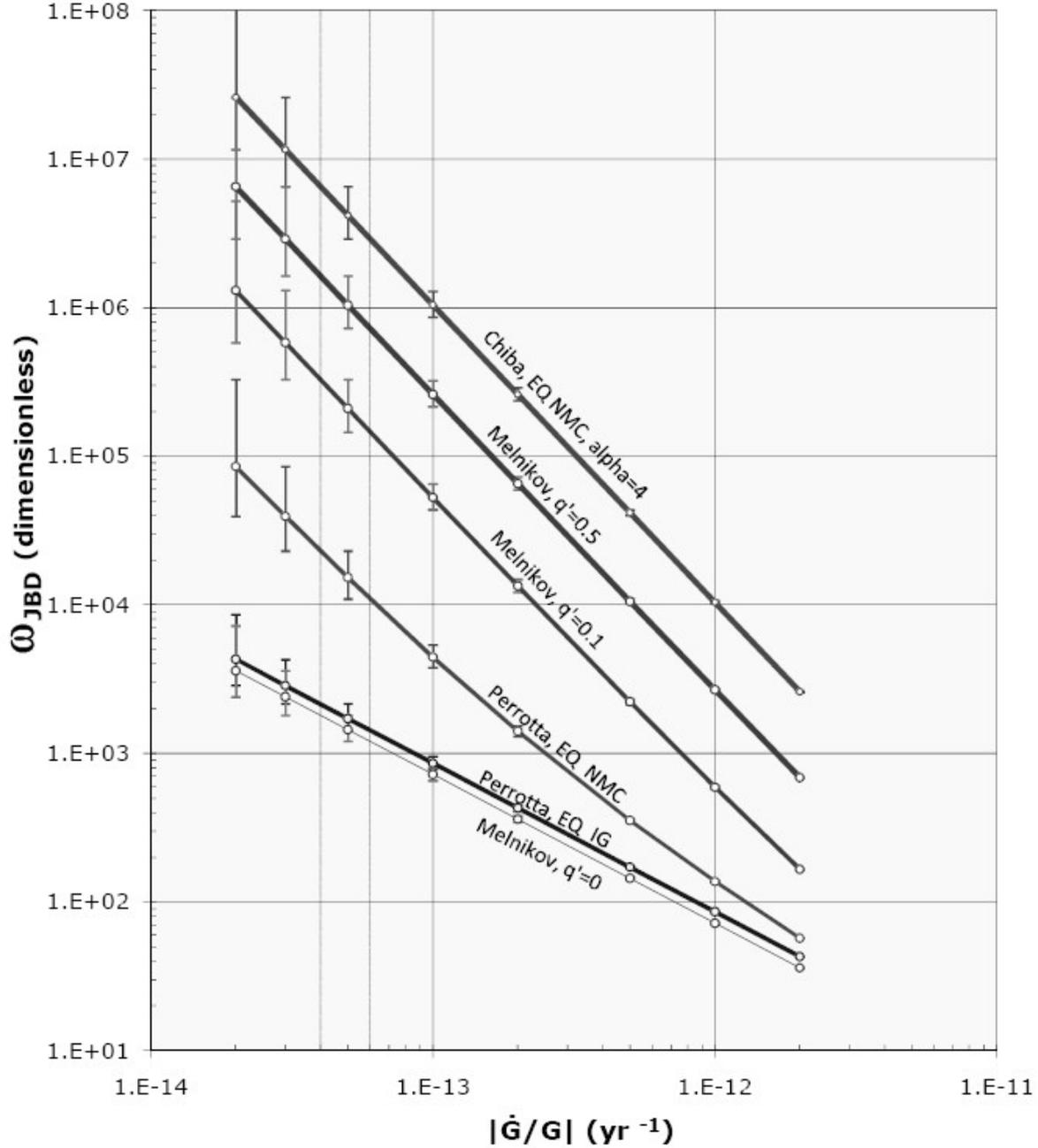

**Fig 3.** The predicted value of $\omega_{JBD}$ varies strongly with $|\dot{G}/G|$ and is very model-dependent among the various scalar-tensor theories. Note especially that the two Non-Minimal Coupling (NMC) models yield very different curves. The terminology is as follows. Chiba: Extended Quintessence (EQ) with Non-Minimal Coupling (NMC); Perrotta *et al.*: two EQ models with Induced Gravity (IG) and Non-Minimal Coupling (NMC); Melnikov *et al.*: three scalar-tensor models with selected values of shifted deceleration parameter, $q'$. The error bars shown for $\omega_{JBD}$ are illustrative and are based on the assumption that $\Delta|\dot{G}/G| = 1 \times 10^{-14}$ yr$^{-1}$ in all cases. For illustration, the region of $|\dot{G}/G|$ between 4 and 6 x $10^{-14}$ is marked by vertical lines (see text).